\documentclass[aps,showpacs,showkeys,prl,preprintnumbers,nofootinbib]{revtex4}
\usepackage{epsfig}

\usepackage{color}
\usepackage{graphicx}
\usepackage{bbm}

\newcommand{\nc}{\newcommand}
\nc{\non}{\nonumber}
\nc{\hc}{\hbox {H.c.}} 
\nc{\noi}{\noindent}
\nc{\barx}{\bar{x}}
\nc{\pbarn}{\;\hbox {pb}}
\nc{\fbarn}{\;\hbox {fb}}
\nc{\lsp}{\;\;\;\;\;}
\nc{\Lsp}{\;\;\;\;\;\;\;\;\;\;}  
\nc{\LLsp}{\lspace \lspace}
\nc{\lra}{\longrightarrow}
\nc{\beq}{\begin{equation}}  \nc{\eeq}{\end{equation}}
\nc{\bea}{\begin{eqnarray}}  \nc{\eea}{\end{eqnarray}}
\nc{\baa}{\begin{array}}     \nc{\eaa}{\end{array}}
\nc{\bit}{\begin{itemize}}   \nc{\eit}{\end{itemize}}
\nc{\ben}{\begin{enumerate}} \nc{\een}{\end{enumerate}}
\nc{\bce}{\begin{center}}    \nc{\ece}{\end{center}}
\nc{\bpm}{\begin{pmatrix}}   \nc{\epm}{\end{pmatrix}}
\nc{\bvt}{\begin{verbatim}}  \nc{\evt}{\end{verbatim}}

\def\gesim{\,{\raise-3pt\hbox{$\sim$}}\!\!\!\!\!{\raise2pt\hbox{$>$}}\,}
\def\lesim{\,{\raise-3pt\hbox{$\sim$}}\!\!\!\!\!{\raise2pt\hbox{$<$}}\,}

\def\gev{\;\hbox{GeV}}
\def\tev{\;\hbox{TeV}}

\def\m{\;\hbox{m}}

\def\lsp{\qquad}
\def\lsim{\lesim}
\def\gsim{\gesim}
\def\hc{\hbox{H.c.}}

\def\vev{vacuum expectation value}
\def\lcal{{\cal L}}
\def\ocal{{\cal O}}

\def\mati{{\mathbbm1}}

\def\vevof#1{\left\langle#1\right\rangle}

\nc{\Lam}{\Lambda}
\nc{\Lams}{\Lambda^2}
\nc{\mhs}{m_h^2}
\nc{\mws}{m_W^2}
\nc{\mzs}{m_Z^2}
\nc{\mts}{m_t^2}
\nc{\mh}{m_h}
\nc{\mw}{m_W}
\nc{\mz}{m_Z}
\nc{\mt}{m_t}
\nc{\vp}{\varphi}
\nc{\mpl}{m_{Pl}}

\nc{\lamp}{\lambda_H}
\nc{\lamvp}{\lambda_\varphi}
\nc{\lamx}{\lambda_x}
\nc{\xf}{x_f}

\begin{document}
\preprint{IFT-09-11 \cr UCRHEP-T476}

\title{Na\"ive solution of the little hierarchy problem \\
and its physical consequences\footnote{Presented at the XXXIII International Conference of Theoretical Physics,
"Matter to the Deepest: Recent Developments in Physics
of Fundamental Interactions", Ustro\'n 2009.}
}
\author{Bohdan GRZADKOWSKI}
\email{bohdan.grzadkowski@fuw.edu.pl}
\affiliation{Institute of Theoretical Physics,  University of Warsaw,
Ho\.za 69, PL-00-681 Warsaw, Poland}

\author{Jos\'e WUDKA}
\email{jose.wudka@ucr.edu}
\affiliation{Department of Physics, University of California,
Riverside CA 92521-0413, USA}

\begin{abstract}
We argue that adding gauge-singlet real scalars to
the Standard Model can both ameliorate the little hierarchy problem
and provide a realistic source of Dark Matter. Masses of the scalars
should be in the $1-3\tev$ range, while the lowest cutoff of
the (unspecified) UV completion of the model must be $ \gesim 5\tev$, 
depending on the
Higgs boson mass and the number of singlets present. 
The scalars couple to the Majorana mass term for right-handed
neutrinos implying one massless neutrino.  
The resulting mixing angles are consistent with the 
tri-bimaximal mixing scenario.
\end{abstract}


\maketitle
  
\section{Introduction}

Our intention is to construct economic extension of
the Standard Model (SM) for which the little hierarchy problem is
ameliorated while preserving all the successes of the SM.  
We will consider
only those extensions that  interact with the SM through
renormalizable interactions. 
Since quadratic divergences of the Higgs boson mass are dominated by
top-quark contributions, it
is natural to consider extensions of the scalar sector, so that
they can reduce the top contribution (as they enter with an opposite sign).
The extensions we consider, although renormalizable, shall
be treated as effective low-energy theories valid
below a cutoff energy $\sim 5-10 \tev$; we will not discuss the UV
completion of this model.

\section{The little hierarchy problem}
The quadratically divergent 1-loop correction to the
Higgs boson ($h$) mass was first calculated by Veltman~\cite{Veltman:1980mj}
\beq
\delta^{(SM)} \mhs =\left[
3\mts/2-(6\mws+3\mzs)/8  - 3\mhs/8  \right]  \Lams/(\pi^2 v^2)
\label{hcor}
\eeq
where $ \Lam$ is a UV cutoff, that we adopt as a regulator,
and $v \simeq 246 \gev $ denotes the \vev\ of the scalar doublet
(SM logarithmic corrections are small since we assume $v \ll \Lam \lesim
10 \tev$). The SM is considered here as an effective theory valid up to
the physical cutoff $\Lam$, the scale at which new
physics enters.

Precision tests of the SM (mainly from the oblique $T_{\rm obl}$
parameter ~\cite{Amsler:2008zz}) require a light Higgs boson, $m_h
\sim 120-170 \gev$. The correction (\ref{hcor}) can then exceed
the mass itself even for small values of $ \Lam $, e.g. 
$\delta^{(SM)} \mh^2 \simeq \mh^2$ for $\mh = 130
\gev$ already for $\Lam
\simeq 600 \gev$. That suggests extensions of the SM with a typical
scale at $1\tev$, however no indication of such low energy new
physics have been observed.  This difficulty is known as the little
hierarchy problem.

Here our modest goal is to
construct a simple modification of the SM within which $\delta
\mhs$ (the total correction to the SM Higgs boson mass squared) is
suppressed up to only $ \Lam \lsim 3-10 \tev$.  Since (\ref{hcor}) is
dominated by the fermionic (top quark) terms, the most economic way
of achieving this is by introducing new scalars $\vp_i$ whose
1-loop contributions reduce the ones derived from the SM.  In order
to retain SM predictions we assume that $\vp_i$ are singlets
under the SM gauge group. Then it is easy to observe that the theoretical expectations for all existing experimental tests remain unchanged if $\langle \vp_i \rangle =0$ (which we assume hereafter), in particular the SM 
expectation of a light Higgs is
preserved. 

The most general scalar potential implied by $Z_2^{(i)}$ independent
symmetries $\vp_i\to -\vp_i$ (imposed in order to 
prevent $\vp_i \to hh$ decays) reads:
\beq
V(H,\vp_i)=-\mu_H^2 |H|^2 + \lamp |H|^4
+ \sum_{i=1}^{N_\vp}(\mu_\vp^{(i)})^2 \vp_i^2 + 
\frac{1}{24}  \sum_{i,j=1}^{N_\vp} \lambda_\vp^{(ij)} \vp_i^2 \vp_j^2
+ |H|^2 \sum_{i=1}^{N_\vp}\lamx^{(i)}  \vp_i^2
\label{pot}
\eeq
In the following numerical computations we assume for simplicity
that $\mu_\vp^{(i)}=\mu_\vp$, $\lambda_\vp^{(ij)}=\lambda_\vp$ and
$\lamx^{(i)}=\lamx$,
in which case (\ref{pot}) has an $ O(N_\vp)$ symmetry
(small deviations from  this assumption do 
not change our results qualitatively).
The minimum of $V$ is at $\vevof H = v/\sqrt{2}$ and
$\vevof{\vp_i} = 0$ when $ \mu_\vp^2 > 0$ and $\lamx, \lamp > 0$ 
which we now assume. The masses
for the SM Higgs boson and the new scalar singlets are 
$\mhs=2\mu_H^2$ and $m^2
= 2\mu_\vp^2+\lamx v^2$ ($\lamp v^2=\mu_H^2$), respectively.

Positivity of the potential at large field strengths
requires $ \lamp \lambda_\vp > 6 \lamx^2 $ at the tree level.
The high energy unitarity  (known~\cite{Cynolter:2004cq} for $N_\vp=1$) 
implies $\lamp \leq 4\pi/3$ (the SM requirement) and $\lambda_\vp \leq 8 \pi$, $\lamx <
4\pi$. These conditions, however, are derived from the
behavior of the theory at energies $E \gg m$, where we don't pretend
our model to be valid, so that neither the stability limit nor the
unitarity constraints are applicable within our pragmatic
strategy, which 
aims at a modest increase of $ \Lam$
to the $3-10\tev$ range.

\begin{figure}[ht]
\centering \includegraphics[width=6cm]{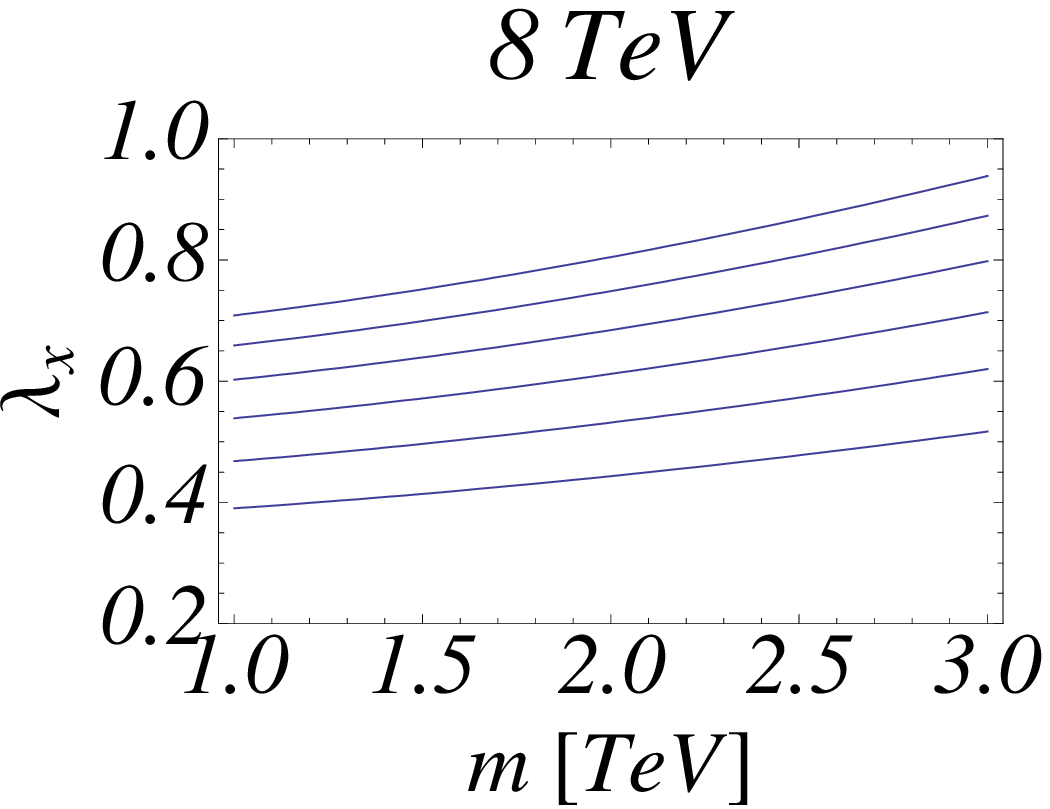}
\centering \includegraphics[width=6cm]{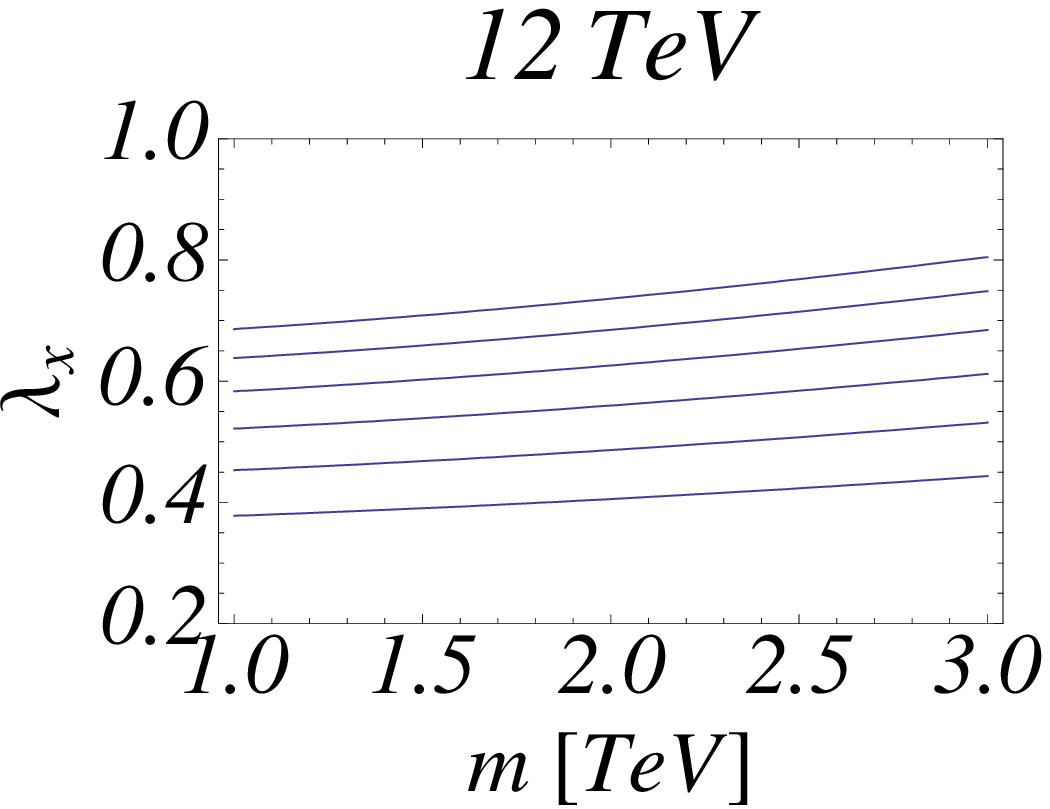}
\caption{Plot of $\lamx$ 
corresponding to $D_t=0$ and $N_\vp=6$
as a function of $m$ for
$\Lam=8\tev$ and $12\tev$ (as indicated above 
each panel). 
The various curves correspond to
$\mh=130, ~150, ~170, ~190, ~210, ~230 \gev$ (starting with
the uppermost curve).
}
\label{lam}
\end{figure}

The existence of $ \vp_i $ generates additional radiative
corrections~\footnote{ The $ \Lam^2 $
corrections to $m^2$ can also be tamed within the full model
with additional fine tuning,
but we will not consider them here, see~\cite{BJ}.}
to $\mhs$. Then the extra
contribution to $\mhs$ reads
\beq
\delta^{(\vp)} \mhs = - [N_\vp\lamx/(8\pi^2)]\left[ \Lam^2 - m^2
\ln\left(1+\Lam^2/m^2\right)\right]
\label{hcorx}
\eeq
Adopting the parameterization $|\delta \mhs|=|\delta^{(SM)} \mhs +
\delta^{(\vp)} \mhs| = D_t \mhs$, we can
determine the value of $\lamx$ needed to suppress $\delta \mhs$ to a
desired level ($D_t$) as a function of $m$, for any choice of $\mh$ and
$\Lam$; examples are plotted in fig.\ref{lam} for $N_\vp=6$.
It should be noted that (in contrast to SUSY) the logarithmic terms in
(\ref{hcorx}) can be relevant in canceling large contributions to
$\delta \mhs$.   
It is important to note that the required value of $ \lamx $ decreases
as  the number of singlets $N_\vp$ grows.   
When $m \ll \Lam$, the $\lamx$ needed for the amelioration of the
hierarchy problem is insensitive to $m$, $D_t$ or $\Lam$; as
illustrated in fig.\ref{lam}; analytically we find up to terms $\ocal \left(m^4/\Lam^4\right)$
\beq 
\lamx \simeq N_\vp^{-1}
\left\{4.8 - 3 (\mh/v)^2 + 2D_t
[2\pi/(\Lam/\tev)]^2\right\}
\left[1-m^2/\Lam^2\ln \left(m^2/\Lam^2\right)\right]\,.
\label{laxaprox}
\eeq 
Since we consider $\lamx \sim \ocal(1) $
effects of higher order corrections~\cite{Einhorn:1992um}  to 
(\ref{hcor}) should be considered as well (see also \cite{Kolda:2000wi}).  
In general, the fine 
tunning condition reads
($\mh$ was chosen as a renormalization scale):
\beq 
|\delta^{(SM)} \mhs + \delta^{(\vp)} \mhs + \Lam^2
\sum_{n=1}f_n(\lamx, \dots)
\left[ \ln(\Lam/\mh) \right]^n |
= D_t \mhs\,,
\label{hor}
\eeq 
where the coefficients $f_n(\lamx, \dots)$ can be determined
recursively~\cite{Einhorn:1992um}, with the leading
contributions being generated by loops containing powers of $\lamx$:
$f_n(\lamx, \dots)\sim [\lamx/(16 \pi^2)]^{n+1}$. To estimate
these effects we can consider the case where 
$\delta^{(SM)} \mhs + \delta^{(\vp)} \mhs = 0$ at one loop then,
keeping only
terms $\propto \lamx^2 $,  we find (using~\cite{Einhorn:1992um}), at 2 loops, 
$D_t\simeq  (\Lam/(4\pi^2 m_h))^2 \; \ln(\Lam/\mh)$ 
(note that $N_\vp \lamx \simeq 4$).
Requiring $D_t \lsim 1$ implies 
$\Lam \lsim  3-5\tev$ for $\mh=130-230\gev$.

It must be emphasized that in the model proposed here
the hierarchy problem is softened (by lifting the cutoff) only if
$\lamx$, $ \Lam $ and $m$ are appropriately fine-tuned;
this fine tuning, however, is significantly less 
dramatic than in the SM. In order to illustrate
the necessary amount of tunning, 
it is useful to calculate
the Barbieri-Giudice~\cite{Barbieri:1987fn} parameter
\beq
\Delta_{\lamx} \equiv (\lamx/\mhs)(\partial \mhs/\partial \lamx) =
|\delta^{(\vp)}\mhs|/\mhs 
\eeq
It turns out that the minimal value of $\Delta_{\lamx}$ 
obtained while scanning over $\lamx,\Lam$ and $m$ 
($0.2 \leq \lamx \leq 6$, $1\tev \leq m \leq 10\tev$ and 
$10\tev \leq \Lam \leq 20\tev$) is substantial: $\Delta_{\lamx} \gsim 200$.


\section{Dark matter}
The singlets $\vp_i$ also offer a natural source for Dark Matter (DM) 
(for  $N_\vp=1$ see \cite{Silveira:1985ke}).  
Using standard techniques for cold DM~\cite{Kolb:1990vq} 
we estimate its present abundance $\Omega_{DM}$,
assuming for simplicity that all the $ \vp_i $
are equally abundant (e.g. as in the $O(N_\vp)$ limit). 
$\Omega_{DM}$ is determined by the thermally  
averaged cross-section for  $\vp_i $
annihilation into SM final states  $\vp_i \vp_i \to SM~SM$,
which 
in the non-relativistic approximation, and for $ m \gg \mh $,
reads
\beq
\left\langle \sigma_i v \right\rangle
\simeq 
 \lamx^2/(8 \pi m^2)
+ \lamx^2 v^2 \Gamma_h(2 m)/(8 m^5)
 \simeq [1.73/(8\pi)] \; \lamx^2/m^2
\label{sigv}
\end{equation}
The first contribution in (\ref{sigv}) originates from the $hh$ final state 
(keeping
only the s-channel Higgs exchange; the  t and u channels can be neglected
since $m \gg \mh$) while the second one comes from all other
final states; $\Gamma_h(2m) \simeq 0.48 \tev ( 2m/ 1 \tev)^3$ 
is the Higgs boson width
calculated for its mass equal $2 m$.

\begin{figure}[ht]
\centering
\includegraphics[width=6cm]{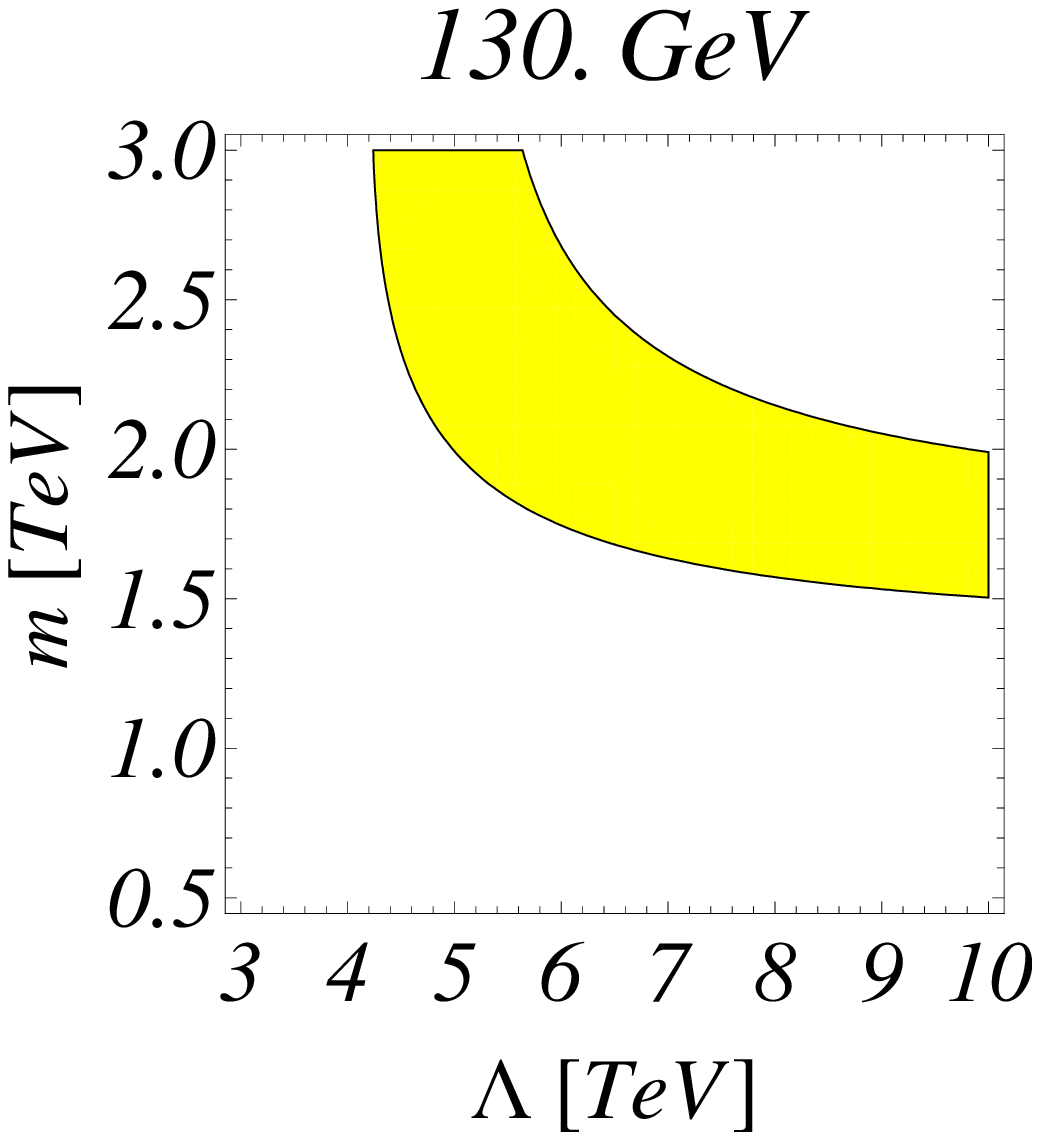}
\includegraphics[width=6cm]{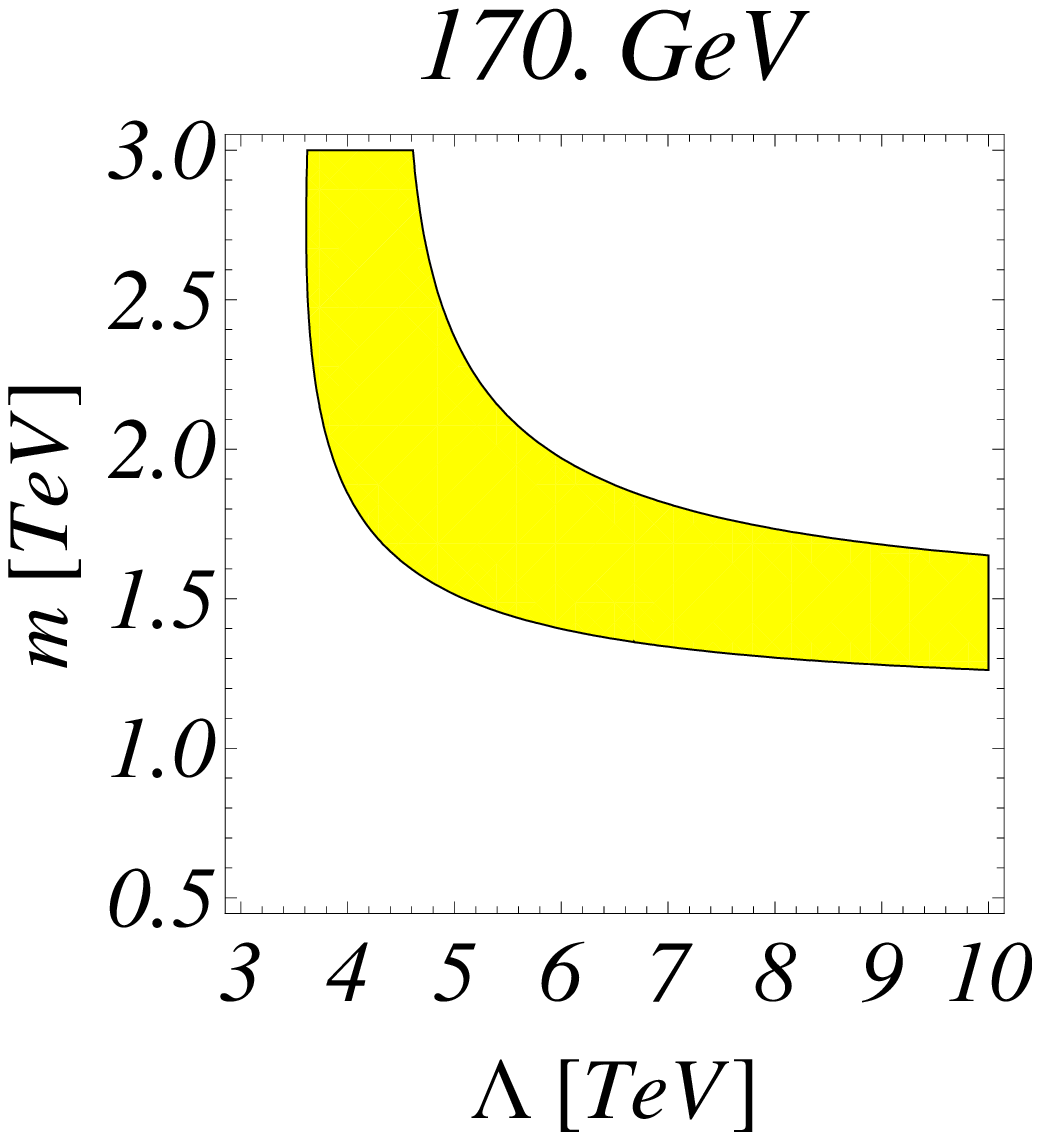}
\caption{The allowed region in the $(m,\Lam)$ plane for $ D_t = 0$, $N_\vp=6$
and $\sum_{i=1}^{N_\vp}\Omega_\vp^{(i)} h^2= 0.106
\pm 0.008$ at the $3\sigma$ level for $\mh=130\gev$ and $170\gev$ (as indicated above 
each panel).}
\label{mLamplot}
\end{figure}

From this the freeze-out temperature $x_f=m/T_f$ is given by
\beq 
x_f=\ln\left[0.038\;
\mpl \; m \vevof{\sigma_i v}/(g_\star x_f)^{1/2} \right]
\eeq
where  $g_\star$ is the number of relativistic degrees of 
freedom 
at annihilation and $\mpl$ denotes the Planck mass. In the 
range of parameters relevant for our purposes, $x_f \sim 12-50$ 
and $m \sim 1-2 \tev$, so that
this is indeed a case of cold DM. 
Then the  present density of $\vp_i$ is
given by
\beq
\Omega_\vp^{(i)} h^2 = 1.06\cdot 10^9 x_f/(g_\star^{1/2} \mpl \langle
\sigma_i v \rangle \gev)\,.
\label{om}
\eeq
The condition that the $\vp_i $'s account for
the observed DM abundance,
$\Omega_{DM}h^2=\sum_{i=1}^{N_\vp}\Omega_\vp^{(i)} h^2= 0.106
\pm 0.008$~\cite{Amsler:2008zz},
can be used to fix $\langle \sigma_i v \rangle$, 
which implies a relation  $\lamx = \lamx(m)$
through (\ref{sigv}).
Using this in the condition  $|\delta \mhs|=D_t \mhs$, we find a relation
between $m$ and $\Lam$ (for a given $D_t$), which is plotted in
fig.\ref{mLamplot} for $N_\vp=6$. 
It should be emphasized that it turns out to be possible to find 
$\Lam$, $\lamx$ and $m$ such that {\it both} the hierarchy is
ameliorated to the desired level {\it and} such that $\Omega_\vp
h^2$ agrees with the DM requirement (we use a $3\sigma$ interval).
It is also instructive to mention  that the singlet
mass (as required by the DM) scales with their multiplicity as $N_\vp^{-3/2}$, therefore growing 
$N_\vp$ implies smaller scalar mass, e.g. changing $N_\vp$ from 1 to 6
leads to the  reduction of mass by a factor $\sim 15$.

%

\section{Neutrino masses and mixing angles}

We now consider implications of the existence of $\vp_i$ for the leptonic
sector, which we assume consists of the SM fields and three
right-handed neutrino fields $\nu_{i\,R}$ ($i=1,2,3$) that are also gauge
singlets. For simplicity here we limit ourself to the case of  only one singlet. 
The relevant Lagrangian is then
\beq
\lcal_Y = -\bar L Y_l H l_R - \bar L Y_\nu \tilde H \nu_R -
\frac12 \overline{(\nu_R)^c} M \nu_R - \vp \overline{(\nu_R)^c}
Y_\vp \nu_R + \hc
\label{lyuk}
\eeq 
where $L=(\nu_L,l_L)^T$ is a SM lepton $SU(2)$ doublet  and $l_R$ 
a charged lepton singlet (we omit family indices); we will assume that the
see-saw mechanism is responsible for smallness of three light neutrino
masses, and therefore we require $M\gg M_D \equiv Y_\nu
v/\sqrt{2}$. Since the symmetry of the potential under $\vp \to -\vp$ should be
extended to (\ref{lyuk}) we require
\beq
L \to S_L L, \; l_R \to S_{l_R} l_R, \; \nu_R \to S_{\nu_R} \nu_R 
\label{trans}
\eeq 
where the unitary matrices $ S_{L, l_R, \nu_R} $ obey
\beq
\quad S_L^\dagger Y_l S_{l_R} = Y_l, \quad \quad S_L^\dagger Y_\nu
S_{\nu_R} = Y_\nu, \quad S_{\nu_R}^T M S_{\nu_R} = + M, \quad
S_{\nu_R}^T Y_\vp S_{\nu_R} = - Y_\vp
\label{inv}
\eeq

In the following we will adopt a basis in which $M$ and $Y_l$ are real and
diagonal; for simplicity we will also assume that $M$ has no
degenerate eigenvalues.  Then the last two conditions in (\ref{inv})
imply that $S_{\nu_R}$ is real and diagonal (so $\pm 1$).  
It is easy to see that for $3$ neutrino species there are two 
possibilities (up to
permutations of the basis vectors): we either have 
$ S_{\nu_R} = \pm \mati,~ Y_\vp =0 $, or, more interestingly,
\beq
S_{\nu_R} = \epsilon \; \hbox{diag}(1,1,-1); \quad
Y_\vp = \bpm{0 & 0 & b_1 \cr 0 & 0 & b_2 \cr b_1 & b_2 & 0 }\epm ,
\qquad \epsilon = \pm1\,,
\label{snur.sol}
\eeq
where $ b_{1,2} $ are, in general,
complex. To satisfy the first  conditions in (\ref{inv}) one needs 
$S_{l_R}=S_L$ with
\beq
S_L = \hbox{diag}(s_1, s_2,s_3) , \quad |s_i| = 1 
\label{SL}
\eeq
Diagonalizing (to leading order in $M^{-1}$) the neutrino mass matrix
in terms of the light ($n$) and heavy ($N$) eigenstates leads to
\beq
\lcal_m=-(\bar n M_n n + \bar N M N/2) \; {\rm with } \;
 M_n = \mu^* P_R + \mu P_L, ~ \mu = 
- 4 M_D M^{-1} M_D^T
\eeq
where $n$ and $N$ are related to $\nu_R$ and $\nu_L$ through $\nu_L =
n_L + (M_D M^{-1}) N_L$ and $\nu_R = N_R - (M^{-1} M_D^T)
n_R$. 

The remaining condition in (\ref{inv}) leads to ten  inequivalent solutions 
for $Y_\nu$~\footnote{The conditions (\ref{inv}) where
also discussed in \cite{Low:2005yc}.}.  
Of those, assuming no more than one massless neutrino and the 
absence of $\vp\to n_i n_j$ decays, only one is interesting;
it  corresponds to $ s_{1,2,3} = \epsilon $ (see (\ref{snur.sol})).
Since $\det Y_\nu = 0$, the symmetry implies
one massless neutrino. 

To compare our results with experimental constraints on the leptonic mixing
angles, we use the so-called 
tri-bimaximal~\cite{Harrison:2002er} lepton mixing matrix where
$\theta_{13}=0$, $\theta_{23}=\pi/4$ 
and $\theta_{12}=\arcsin(1/\sqrt{3})$. We find that the
form of $Y_\nu$ consistent with this (up to axes permutations) is
\bea
Y_\nu &=& \left(\baa{ccc}a&b&0\\-a/2&b&0\\-a/2&b&0\eaa\right)\,,
\baa{l}m_1=-3v^2a^2/M_1\\m_2=-6v^2b^2/M_2\\m_3=0\eaa 
\label{ynu}
\eea
where $a$ and $b$ are real (for simplicity) parameters.
The resulting mass spectrum agrees 
with the observed pattern of neutrino mass differences, see
e.g. \cite{Altarelli:2007gb}.
If $\vp$ is a candidate for DM we should guarantee its stability. 
For the solution (\ref{snur.sol}) only $N_3$ and $ \vp $
are odd under the $Z_2$ symmetry hence the  $\vp$ will
be absolutely stable if $m<M_3$.

It is worth noticing that in the presence of $ Y_\vp $ 
there exist three sources of 1-loop contributions to the quadratic divergence 
in corrections to the $\vp$ mass: {\it 1)} those generated by 
$|H|^2 \sum_{i=1}^{N_\vp}\lamx^{(i)}  \vp_i^2$, {\it  2)} those
from the quartic $\vp$ coupling and {\it  3)}
the additional one from the Yukawa coupling 
$\vp \overline{(\nu_R)^c} Y_\vp \nu_R$. The presence of  $\nu_R$ can be
used~\cite{BJ} to ameliorate the little hierarchy problem associated with $ m $
thereby
``closing'' the solution to the little hierarchy problem in a spirit
analogous to supersymmetry.

\section{Conclusions}
We have shown that the addition of real scalar singlets $\vp_i$ to the SM may
soften  the little hierarchy problem (by lifting the cutoff $\Lam$
to multi TeV range). This scenario also offers realistic candidates for DM. In
the presence of right-handed neutrinos this model allows for
a light neutrino mass matrix texture that is consistent
with experimental data, in which case there should be 
one massless neutrino.

\bce {\bf Acknowledgments} \ece

We thank the Organizers of  
the XXXIII International Conference of Theoretical Physics
for their warm hospitality during the meeting.
This work is supported in part by the Ministry of Science and Higher
Education (Poland) as research project N~N202~006334 (2008-11) 
and by the U.S. Department of Energy
grant No.~DE-FG03-94ER40837.  B.G. acknowledges support of the
European Community within the Marie Curie Research \& Training
Networks: "HEPTOOLS" (MRTN-CT-2006-035505), and "UniverseNet"
(MRTN-CT-2006-035863), and through the Marie Curie Host Fellowships
for the Transfer of Knowledge Project MTKD-CT-2005-029466.


\begin{thebibliography}{99}

\bibitem{Veltman:1980mj}
  M.~J.~G.~Veltman,
  Acta Phys.\ Polon.\  B {\bf 12}, 437 (1981).

\bibitem{Amsler:2008zz}
  C.~Amsler {\it et al.}  [Particle Data Group],
  Phys.\ Lett.\  B {\bf 667}, 1 (2008).

\bibitem{Cynolter:2004cq}
  G.~Cynolter, E.~Lendvai and G.~Pocsik,
  Acta Phys.\ Polon.\  B {\bf 36}, 827 (2005)
  [hep-ph/0410102].

\bibitem{BJ}
B.~Grzadkowski and J.~Wudka, work in progress.
 
\bibitem{Einhorn:1992um}
  M.~B.~Einhorn and D.~R.~T.~Jones,
  Phys.\ Rev.\  D {\bf 46}, 5206 (1992).
  
\bibitem{Kolda:2000wi}
  C.~F.~Kolda and H.~Murayama,
  JHEP {\bf 0007}, 035 (2000)
  [hep-ph/0003170];
  J.~A.~Casas, J.~R.~Espinosa and I.~Hidalgo,
  JHEP {\bf 0411}, 057 (2004)
  [hep-ph/0410298].

\bibitem{Barbieri:1987fn}
  R.~Barbieri and G.~F.~Giudice,
  Nucl.\ Phys.\  B {\bf 306}, 63 (1988).

\bibitem{Silveira:1985ke}
  V.~Silveira and A.~Zee,
  Phys.\ Lett.\  B {\bf 157}, 191 (1985);
  J.~McDonald,
  Phys.\ Rev.\  D {\bf 50}, 3637 (1994)
  [hep-ph/0702143];
  C.~P.~Burgess, M.~Pospelov and T.~ter Veldhuis,
  Nucl.\ Phys.\  B {\bf 619}, 709 (2001)
  [hep-ph/0011335];
  H.~Davoudiasl, R.~Kitano, T.~Li and H.~Murayama,
  Phys.\ Lett.\  B {\bf 609}, 117 (2005)
  [hep-ph/0405097];
  J.~J.~van der Bij,
  Phys.\ Lett.\  B {\bf 636}, 56 (2006)
  [hep-ph/0603082];
  O.~Bertolami and R.~Rosenfeld,
  [arXiv:0708.1784];
  V.~Barger, P.~Langacker, M.~McCaskey, M.~J.~Ramsey-Musolf and G.~Shaughnessy,
  Phys.\ Rev.\  D {\bf 77}, 035005 (2008)
  [arXiv:0706.4311];
  S.~Andreas, T.~Hambye and M.~H.~G.~Tytgat,
  JCAP {\bf 0810}, 034 (2008)
  [arXiv:0808.0255].
 
\bibitem{Kolb:1990vq}
  E.~W.~Kolb and M.~S.~Turner,
  Front.\ Phys.\  {\bf 69} (1990) 1.

\bibitem{Low:2005yc}
  C.~I.~Low,
  Phys.\ Rev.\  D {\bf 71}, 073007 (2005)
  [hep-ph/0501251];
  S.~S.~C.~Law and R.~R.~Volkas,
  Phys.\ Rev.\  D {\bf 75}, 043510 (2007)
  [hep-ph/0701189].
 
\bibitem{Harrison:2002er}
  P.~F.~Harrison, D.~H.~Perkins and W.~G.~Scott,
  Phys.\ Lett.\  B {\bf 530}, 167 (2002)
  [hep-ph/0202074].
  
\bibitem{Altarelli:2007gb}
  G.~Altarelli,
  [arXiv:0711.0161].

\end{thebibliography}
\end{document}